\documentclass[10pt]{iopart}
\usepackage{iopams}
\usepackage{graphicx}% Include figure files
\usepackage{cite}
\begin{document}

%\title[Short title]{Long title}
\title[Temperature dependence of thermal jamming]{Temperature dependence of the transition packing fraction of thermal jamming in a harmonic soft sphere system}

\author{Moumita Maiti$^1$}
\address{$^1$ Institut f\"ur Physikalische Chemie, Westf\"alische Wilhelms-Universit\"at (WWU), Corrensstr. 28/30, 48149 Münster, Germany}
\author{Michael Schmiedeberg$^{2}$}
\address{$^2$ Institut f\"ur Theoretische Physik I, Friedrich-Alexander Universit\"at Erlangen-N\"urnberg (FAU), Staudtstra{\ss}e 7, 91058 Erlangen, Germany}
%\ead{Email}							%one person's email
\eads{\mailto{michael.schmiedeberg@fau.de}}	%more emails
\vspace{10pt}
\begin{indented}
\item[]September 2018
\end{indented}

\begin{abstract}
  The glassy dynamics of soft harmonic spheres is often mapped onto the dynamics of hard spheres by considering an effective diameter for the soft particles and therefore an effective packing fraction. While in this approach the thermal fluctuations within valleys of the energy landscape are covered, the crossing of energy barriers from one valley into another usually is neglected. Here we argue - motivated by studies of the glass transition based on explorations of the energy landscape - that the crossing of energy barriers can be attributed by an effective decrease of the glass transition packing fraction with increasing temperature $T$ according to $T^{0.2}$. Furthermore, we reanalyzing data of soft sphere simulations. Since fitting scaling laws to simulation data always allows for some arbitrariness, we cannot prove based on the simulation data that our idea of a shift of the glass transition packing fraction due to barrier crossings is the only possible way to explain the discrepancies that have been observed previously. However, we show that a possible explanation of the simulation data is given by our approach to characterize the dynamics of soft spheres by both, the previously-considered temperature-dependent effective packing fraction due to the increase of the mean overlap between neighboring particles with stronger thermal fluctuations and the newly introduced increase of the glass transition packing with an increasing probability of barrier crossings.
\end{abstract}

%\pacs{64.70.kj,82.70.Dd, 64.70.qd}

\vspace{2pc}

%\submitto{\JPCM}

This is the version of the article before peer review or editing, as submitted by an author to Journal of Physics: Condensed Matter. IOP Publishing Ltd is not responsible for any errors or omissions in this version of the manuscript or any version derived from it. The Version of Record is available online at https://doi.org/10.1088/1361-648X/ab01e9.\\
Citation of the final version: Moumita Maiti and Michael Schmiedeberg 2019 J. Phys.: Condens. Matter 31 165101\\

\maketitle
\ioptwocol
\section{\label{sec:introduction}Introduction}
Upon decrease of the temperature or increase of the density many particulate systems exhibit a dramatic slowdown of the dynamics. At the so-called glass transitions the relaxation dynamics becomes longer than the typical duration of an experiment or simulations (see, e.g. \cite{Woodcock,PuseyandMegen,Angell,BerthierandBiroli}). A simple and widely-used model system to explore the glassy dynamics or the transition where the system jams consists of soft spheres with harmonic interactions, i.e., particles that do not interact if they do not overlap and otherwise repel each other with a force proportional to the overlap. Note that in this article we are interested in the dynamical glass transition where the dynamics becomes slower than the typical timescale of observation. Slow ageing dynamics might still occur and therefore some properties might be history dependent \cite{BerthierandBiroli}. Note that in this article we are not studying any ideal glass transition or Kauzmann temperature \cite{Kauzmann} and we are not considering any speed up of relaxation processes by particle exchange methods \cite{berthier-prx} that recently have been introduced in order to enable the relaxation in system at packing fractions that are above what we consider the glass transition here.

Concerning the glass transition of soft harmonic spheres, in \cite{zhangetal} Zhang et al. found in experiments and simulations that the pair distribution function $g(r)$ possesses a pronounced peak close to the transition. Furthermore, Zhang et al. show that $\Delta\phi_v(T)\propto T^{1/2}$ where $\phi_v(T)$ is the packing fraction with the largest peak in $g(r)$ and $\Delta\phi_v(T)=\phi_v(T)-\phi_{g,0}$ with $\phi_{g,0}=\lim_{T\rightarrow 0}\phi_v(T)$. Finally, by using simulation, they discovered that this scaling of $\phi_v(T)$ cannot correctly describe the behavior of a $\phi_{\tau=const}(T)$-curve that is given by a constant relaxation time $\tau$ \cite{zhangetal}. In \cite{zhangetal} it is argued that this discrepancy can be seen as a difference between the line $\phi_v(T)$ where the peak of $g(r)$ as a function of the packing fraction becomes maximal and the dynamical glass transition line $\phi_g(T)$ that one is expected to approach in case of large $\tau$. In other words, there seems to be a discrepancy between the behavior of the structure and of the dynamics close to the glass transition. Note that for a similar system a link between the correlation length based on clusters of fast particles and the $\alpha$-relaxation times could be established \cite{Zhang2016}.

Berthier and Witten in \cite{BerthierandWitten,BerthierandWitten1} showed that the dynamics of soft harmonic spheres at different temperatures $T$ can be approximately compared to each other if they possess the same effective packing fraction $\phi_{eff}(T)$ that is given by $\phi_{eff}(T)=\phi-a T^{\mu/2}$ with $\mu=1.3$. Note that from the overlap of two soft harmonic spheres one expects $\mu=1$ \cite{zhangetal,BerthierandWitten,BerthierandWitten1}. Berthier and Witten argue, that their reported exponent $\mu=1.3$ is intermediate between the expected $\mu=1$ and $\mu=3/2$. The latter exponent is claimed to occur for ballistic, dilute systems\cite{BerthierandWitten}.

Another approach to determine an effective packing fraction was employed in \cite{schmiedeberg2011}, where it is shown that the dynamics of soft spheres can be mapped on the dynamics of hard spheres with the effective packing fraction $\phi_{eff}(T)$ as given by the Andersen-Weeks-Chandler effective diameter \cite{andersen71} that originally was introduced to map the structure of soft spheres onto hard sphere behavior. Such an approach has been used in other works as well \cite{medina2011,medina2012,medina2013} and it has been demonstrated that the dynamics of various soft particulate systems can be mapped onto the hard sphere dynamics as long as the probability for significant or even almost complete overlaps is small \cite{medina2011,medina2012,medina2013}. In \cite{schmiedeberg2011} it was observed that there is a small systematic overestimation of the effective packing fraction especially in case of larger temperatures.

Note that in all of the works mentioned before the effective packing fraction of the system decreases with increasing temperature $T$, i.e., for packing fractions below the glass transition the distance to the transition increases with increasing $T$.

However, in \cite{maiti2018} we have discovered that the packing fraction of the glass transition $\phi_g(T)$ decreases with increasing $T$, i.e., the distance to the glass transition decreases with increasing $T$. Note that in \cite{maiti2018} fluctuations within valleys of the potential energy landscape are neglected. As a consequence, in \cite{maiti2018} we determine corrections to the glass transition due to the thermal crossing of energy barriers but the decrease of $\phi_{eff}(T)$ does not occur because thermal fluctuations around local equilibrium-like configurations are neglected.

In this article we reanalyze the simulation data of harmonic soft sphere systems where both fluctuations around local equilibrium configurations leading to a $\phi_{eff}(T)$ that decreases with increasing $T$ as in\cite{zhangetal,BerthierandWitten,BerthierandWitten1,schmiedeberg2011} as well as crossings of energy barriers that cause an decreasing $\phi_g(T)$ as studied in \cite{maiti2018}. To be specific, we show that the soft sphere data can be explained with an effective packing fraction $\phi_{eff}(T)$ that decreases with increasing $T$ as determined by the Andersen-Weeks-Chandler-method \cite{andersen71,schmiedeberg2011} or alternatively, in case of small $T$, as described by the scaling $\phi_{eff}(T)=\phi-a T^{\mu/2}$ with the actual packing fraction $\phi$ and $\mu=1.0$ in agreement to \cite{zhangetal} and the line of argumentation for small overlaps in \cite{BerthierandWitten,BerthierandWitten1}. In addition, the glass transition packing fraction $\phi_g(T)$ decreases as $\phi_{g,0}-\phi_g(T)\propto T^{0.2}$, which is in agreement to the results of \cite{maiti2018}. Therefore, the fluctuations of the overlaps lead to an decrease of the effective packing fraction $\phi_{eff}(T)$ and in addition the glass transition $\phi_g(T)$ decreases, such that in total the distance $\Delta \phi_{tot}(T)=\phi_g(T)-\phi_{eff}(T)$ from the glass transition changes with temperature as in 
\begin{eqnarray}
  \nonumber
  \Delta \phi_{tot}(T)-\Delta \phi_{tot}(0)
  &=&\left(\phi_g(T)-\phi_{eff}(T)\right)-\left(\phi_{g,0}-\phi\right)\\
\nonumber
&=&\left(\phi_g(T)-\phi_{g,0}\right)-\left(\phi-\phi_{eff}(T)\right)\\
  \nonumber
&\propto& T^{1/2}-bT^{0.2}\label{eq:scaling}
\end{eqnarray}
with a positive constant $b$.

We organize the article as follows: In section \ref{sec:system} we explain our model system in more detail and comment on the simulation methods. In section \ref{sec:results} the results of our analyses of simulation data is presented and discussed before we conclude in section \ref{sec:conclusions}.

\section{\label{sec:system}System}

\subsection{\label{sec:sys_1}Soft spheres with harmonic interactions}

We consider a soft sphere system in three dimensions where the particles repel each other with a force that is proportional to the overlap, i.e., the pair potential for the spheres $i$ and $j$ is given by
\begin{equation}
V(r_{ij})=\begin{cases}{{\epsilon\left(1-\frac{r_{ij}}{\sigma_{ij}}\right)^2/2, r_{ij}<\sigma_{ij},}\atop{ 0, r\geq\sigma_{ij},}}\end{cases}
\end{equation}
where $r_{ij}$ is the distance between the spheres, $\sigma_{ij}$ is their mean diameter, and $\epsilon$ is a positive constant that denotes the strength of the repulsion. 

The temperature $T$ is given in units of $\epsilon/k_B$ in the data from \cite{schmiedeberg2011} that we analyze in section \ref{sec:res_2}. Originally, the temperature $T$ in \cite{BerthierandWitten,BerthierandWitten1} is given in units of $2\epsilon/k_B$, which we change to $\epsilon/k_B$ in our discussion in section \ref{sec:res_3}.

\subsection{\label{sec:sys_2}Simulation details}

In this article we reanalyze simulation data from \cite{schmiedeberg2011} and from \cite{BerthierandWitten,BerthierandWitten1}. In the following, we shortly comment on the methods employed in those works as well as on the quantities that are considered in the respective data sets. Further details are given in the original articles \cite{schmiedeberg2011,BerthierandWitten,BerthierandWitten1}.

In \cite{schmiedeberg2011} we employed molecular dynamics simulations of a monodisperse system at constant temperature $T$ and constant pressure $p$. The timescale that describes rearrangements is given by the time $\tau$ where the mean square displacement reaches the squared diameter, i.e., $\left\langle r^2 (\tau)\right\rangle=\sigma^2$. For the mapping onto the hard sphere dynamics the dimensionless time $\tau^*=\tau\sqrt{p\sigma/m}$ is used.

In \cite{BerthierandWitten,BerthierandWitten1} Berthier and Witten present molecular dynamics simulations at constant temperature and volume. A bidisperse system with a 50:50 mixtures with a ratio of diameters $\sigma_1/\sigma_2=1.4$ is used. Relaxation times $\tau_{\alpha}$ are measured as $\alpha$-relaxation time by considering the decay of the self-part of the intermediate scattering function and usually the dimensionless timescale $\tau_\alpha^*=\sqrt{T\epsilon/m}\tau_\alpha/\sigma_2$ is employed. The data that we reanalyze in section \ref{sec:res_3} has been extracted from figure 2(a) of \cite{BerthierandWitten}.

Note that different ways to measure the timescales are used in \cite{maiti2018}, \cite{schmiedeberg2011}, and \cite{BerthierandWitten,BerthierandWitten1}. However, within each set of data we stick to one definition, namely the definition employed in the original publication. We assume that for the considered systems the scaling works for all definitions if it is based on the divergence that can be obtained by extrapolation.

\subsection{\label{sec:sys_3}Temperature dependence of thermal jamming}

\begin{figure}[htb]
\centering
\includegraphics[width=\linewidth]{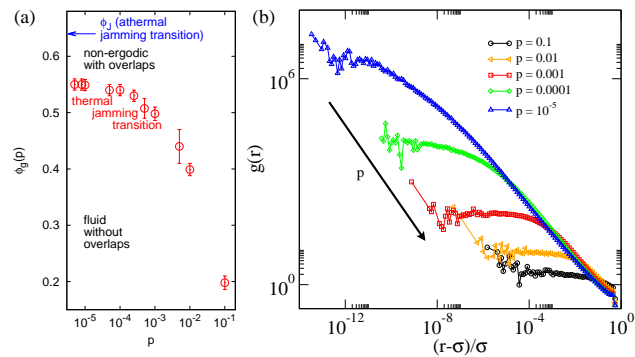}
\caption{(a) Thermal jamming transition packing fractions $\phi_g(p)$ (shown in red) as a function of the probability $p$ for steps where energy barriers can be crossed as determined in \cite{maiti2018}. For packing fractions below the transition line there are no overlaps and the system therefore is unjammed. Above the glass transition packing fraction the overlaps cannot be removed and the only rearrangements are due to the rare barrier crossings (occurring with probability $p$). As a consequence, for $p\rightarrow 0$ the system is effectively non-ergodic, i.e., in the glass phase. The red data points therefore mark the glass transition line for small $p$ (corresponding to small temperatures). The packing fraction of the athermal jamming transition is indicated in blue for comparison. (b) Pair distribution function $g(r)$ for a monodisperse system just above the glass transition determined with the method introduced in \cite{maiti2018} for various probabilities $p$. The $g(r)$ curves along with a discussion of its scaling as a function of $r$ is also contained in the supplemental materials of \cite{maiti2018}.}
\label{fig:glass_line}
\end{figure}

In \cite{maiti2018} we introduced a new approach to explore the energy landscape of a glassy soft sphere system. Motivated by the method employed to determine the athermal jamming transition \cite{OhernLangerLiuNagel,OhernSilbertLiuNagel}, we start with a random initial configuration and then minimize the energy without crossing energy barriers, e.g., by employing a conjugate gradient minimization or a steepest decent approach. Thermal fluctuations around the paths of these minimization methods are not considered. However, in order to denote the possibility that an energy barrier can be crossed, we introduce additional steps that occur with a small probability $p$ and that displace a particle in a random direction such that a barrier can be crossed. Note that we have checked that the details on how these additional steps are implemented do not matter \cite{maiti2018} as long as they enable the rare random crossing of energy barriers. In our simulations with this new method we analyzed whether the ground state can be reached or not. In case the ground state cannot be reached within the time of the simulation the system is effectively non-ergodic. If the ground state is reached, the system is considered to be unjammed \cite{OhernLangerLiuNagel,OhernSilbertLiuNagel}. Another approach to explore and characterize the energy landscape of glasses were employed in \cite{ozawa2012}, where the inherent structures for equilibrated systems are determined and a sharp increase of the obtained jamming packing fraction is observed when the equilibrated systems enter the glass regime \cite{ozawa2012}. Note that our approach in \cite{maiti2018} does not rely on determining inherent structures.

We have confirmed that within our approach the transition packing fraction as a function of the probability $p$ (cf. figure 1(a)) in the limit of $p\rightarrow 0$ does not approach the packing fraction $\phi_J\approx 0.64$ \cite{OhernLangerLiuNagel,OhernSilbertLiuNagel} of athermal jamming that one obtains for $p=0$ \cite{MilzandSchmiedeberg}. Instead the glass transition line $\phi_g(p)$ for small but non-zero $p$ saturates at a lower value $\phi_{g,0}\approx 0.55$. The transition packing fraction $\phi_g(p)$ decreases with increasing $p$, i.e., if energy barriers are crossed more often, the relaxation of the system becomes more difficult. In \cite{maiti2018} we showed that these properties can be explained by the number of particles that are affected by a random barrier crossing. Close to the packing fraction $\phi_g(p)$ a crossing of a barrier can affect almost all particles of the system due to a spatial percolation of the system that also has been observed in \cite{corwin}. Therefore, above $\phi_g(p)$ almost all particles in the system have to restart their relaxation process after a barrier crossing event and as a consequence their relaxation is significantly delayed. Note that a similar mechanism of correlated relaxation processes is also observed in \cite{pastore2015} and percolation transitions have been connected to the glass transition in other systems as well \cite{pastore2013,candia2016}.

Note that in \cite{maiti2018} we also showed that in case of non-instantaneous quenches or in smaller systems glass transition packing fractions above 0.55 are usually observed though the critical behavior remains unchanged. Therefore, in this article we will concentrate on how the glass transition is shifted due to changes of the temperature but not on the absolute values of the dynamical glass transition because the latter obviously are protocol dependent (see also \cite{BerthierandBiroli,berthier-prx,maiti2018}).

\section{\label{sec:results}Results}

\subsection{\label{sec:res_1}Scaling of the peak values of $g(r)$}

\begin{figure}[htb]
\centering
\includegraphics[width=\linewidth]{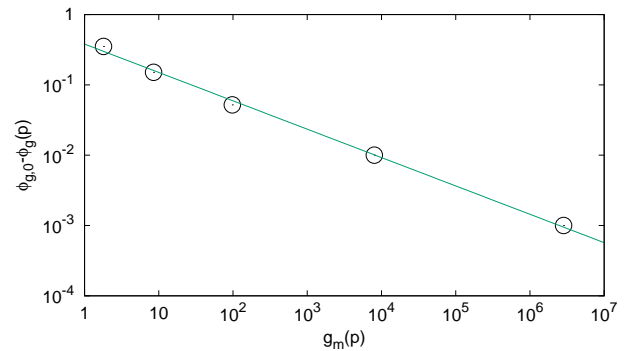}
\caption{Double logarithmic plot of the peak values $g_{m}(p)$ of the pair distribution functions close to the glass transition as a function  $\phi_{g,0}-\phi_g(p)$ where $\phi_{g,0}=0.55$ is the packing fraction of the glass transition in the limit $p\rightarrow 0$ and $\phi_g$ is the packing fraction of the glass transition for a given probability $p>0$. The line is a power law fit according to $g_{m}\propto (\phi_{g,0}-\phi_g(p))^{-1/(2\beta)}$ with $\beta=0.202\pm 0.005$. The data points are determined for the $g(r)$ in figure \ref{fig:glass_line}(b) for the data obtained in \cite{maiti2018}.}
\label{fig:gmax}
\end{figure}

For the case that fluctuations within the valleys of the energy landscape are neglected, we determined how the glass transition packing fraction $\phi_g$ depends on the probability $p$ that determines how often energy barriers are crossed \cite{maiti2018} (see also figure \ref{fig:glass_line}(a)). Obviously the probability $p$ is related to the temperature $T$ of a soft sphere system. In principle, the probability can be approximated by Kramer's rate \cite{haenggi}. However this requires an extensive study of the energy barriers heights as well as the curvatures of the non-trivial energy landscape. Here we employ an alternative approach in order to relate the probability $p$ to the temperature $T$ based on a comparison of properties of the pair distribution functions $g(r)$ close to the glass transition.

For the pair distribution functions $g(r)$ close to the glass transition we consider the peak heights $g_{m}$. The pair distribution functions $g(r)$ have been determined in \cite{maiti2018} for various probabilities $p$ (see also figure \ref{fig:glass_line}(b)). As a result, the peak heights $g_{m}(p)$ are available as a function of $p$. By inverting the the glass transition line $\phi_g(p)$ shown in figure \ref{fig:glass_line}(a) we can express $g_{m}(p)$ as a function of the packing fraction $\phi_g$ (see figure \ref{fig:gmax}). We find that $g_{m}(\phi_g)$ as a function of how much the glass transition packing fraction deviates from the $p\rightarrow 0$ result, i.e., as a function of $\phi_{g,0}-\phi_g(p)$, can be described by a power law $g_{m}\propto (\phi_{g,0}-\phi_g(p))^{-1/(2\beta)}$ with $\beta=0.202\pm 0.005$ (see figure \ref{fig:gmax}). In the following, we will employ the approximation $\beta\approx 0.2$.

According to \cite{zhangetal}, the peak heights $g_{m}$ for soft harmonic spheres depend on the temperature as $g_{m}\propto T^{-1/2}$. As a consequence, we expect $\phi_{g,0}-\phi_g(p)\propto g_{m}^{-2\beta}\propto T^{\beta}$. Therefore, we are able to express the scaling behavior of the glass transition curve $\phi_g(p)$ not only in terms of the probability $p$, but also in terms of the temperature $T$, namely we expect $\phi_{g,0}-\phi_g(T)\propto T^{\beta}$ with $\beta\approx 0.2$. Note that $\phi_g(T)$ is smaller than $\phi_{g,0}$ for $T>0$. 

While fluctuations around a valley of the energy landscape lead to an effective packing fraction that is smaller than the real packing fraction and as a consequence increases the distance from the glass transition for increasing temperature, the crossing of energy barriers can prevent the relaxation and thus leads to a decrease of the glass transition packing fraction as a function of temperature. If the effective packing fraction $\phi_{eff}(T)$ scales like $\phi-\phi_{eff}\propto T^{1/2}$ for harmonic interactions and $\phi_{g,0}-\phi_g(p)\propto T^{\beta}$  with $\beta\approx 0.2$ the distance to the glass line behaves like $\left(\phi_g(T)-\phi_{g,0}\right)-\left(\phi-\phi_{eff}(T)\right)\propto T^{0.5}-bT^{0.2}$ (corresponding to Eq. \ref{eq:scaling}) where $b$ is a positive constant that weights the two scaling contributions. Note that $\phi_{eff}$ can also be determined by calculating effective diameters, e.g., by employing the Andersen-Weeks-Chandler method \cite{andersen71}. Furthermore, our choices to denote the fluctuations around the valley by an decreasing effective diameter instead of an increasing glass transition and to consider the barrier crossing by a deceasing glass transition density instead of an increasing effective packing fraction are arbitrary. The only physical relevant quantity is the difference of the effective packing fraction from the glass line at the respective temperature. This difference is determined such that the soft sphere system behaves similar as a hard sphere system or another soft sphere system with the same difference.

\subsection{\label{sec:res_2}Corrections to the mapping of soft sphere dynamics onto hard sphere dynamics}

In this subsection we reanalyze the simulation data from \cite{schmiedeberg2011}. Our main goal is not not cover a large region of packing fractions and especially we do not aim to get even close to the glass transition packing fraction. For such a task better simulation methods and data are available, especially since Monte Carlo simulations with particle swaps have been developed \cite{berthier-prx}. Our focus here is on the temperature dependence and specifically on the systematic deviations from a hard sphere-like dynamics that have been observed in \cite{schmiedeberg2011} and that are obvious especially for larger temperatures, i.e., temperatures $T\geq 10^{-3}$. In order to quantify the dependence of these deviations on the temperature we employ fits to Vogel-Fulcher-Tammann functions. We want to emphasis that due to the limited range concerning the packing fraction the obtained values of the fit parameters might be somehow arbitrary. However, we are only interested on how these parameters change as a function of temperate. Over the large temperature range of the data from \cite{schmiedeberg2011} these changes are systematic.

\begin{figure}[htb]
\centering
\includegraphics[width=\linewidth]{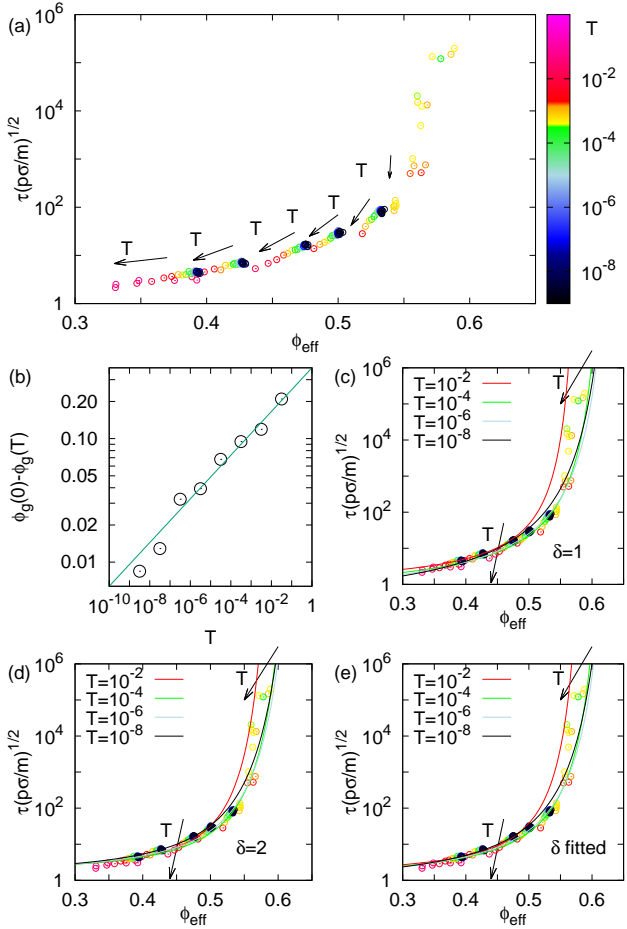}
\caption{(a) Relaxation times as a function of the effective packing fraction. The circles denote the data for harmonic spheres from \cite{schmiedeberg2011}. The trend for increasing temperature is indicated by arrows. The colors indicate the temperature as denoted by the color bar on the right. (b) Glass transition $\phi_{g}(T)$ obtained from Vogel-Fulcher-Tammann extrapolations as a function of temperature $T$. For the extrapolation all data points with a temperature  in an interval $[10.0^{m},10.0^{m+1}]$ with $m=-9,-8,-7,-6,-5,-4,-3,-2$ were used and the corresponding results is shown by one point within each of these temperature intervals. The line is a power law fit. (c-e) Data as in (a) with fitted Vogel-Fulcher-Tammann functions depending on the effective packing fraction and the temperature as described in the text. In (c) the curves diverge with an fixed exponent $\delta=1$. In (d) a fixed $\delta=2$ is used. In (e) $\delta$ is an additional fit parameter. Note that the Vogel-Fulcher-Tammann functions are fitted to all data points but only curves for selected temperatures are shown. As a general trend in all cases for small relaxation times the curves with small temperatures lie above those with low temperature, while for large relaxation times the curves with small temperatures lie below the other curves and are steeper.}
\label{fig:fits_ges}
\end{figure}

In \cite{schmiedeberg2011} the dynamics of soft spheres with an diameter $\sigma$ were compared to the dynamics of hard spheres with a diameter $\sigma_{eff}$, where $\sigma_{eff}$ is the effective diameter as determined by using the Andersen-Weeks-Chandler method \cite{andersen71}. Originally, this method was developed to map the structure of soft sphere systems to the structure of hard spheres with the effective diameter. Concerning the dynamics, it turned out that for small overlaps the soft spheres dynamics can be well-mapped onto the hard sphere dynamics. In figure \ref{fig:fits_ges}(a) the results for small temperatures leading to small overlaps are shown by black and blue circles and can in principle be nicely described by a single curve corresponding to the hard sphere behavior (not shown). However, significant deviations occur for larger overlaps, especially if $\sigma_{eff}/\sigma<0.9$ \cite{schmiedeberg2011}. To be specific, the relaxation times of soft spheres at larger temperatures systematically deviate from the relaxation times of the corresponding hard spheres or data in the $T\rightarrow 0$ limit as can be seen in figure \ref{fig:fits_ges}(a). For small effective packing fractions the relaxation times for larger temperatures (green and red data in figure \ref{fig:fits_ges}(a)) lies below the curve expected from relaxation times that are obtained in the $T\rightarrow 0$ limit (black and blue data in figure \ref{fig:fits_ges}(a)). Note that for larger packing fractions where no obvious trend is visible there might be larger uncertainties concerning the data points (for more details see discussion in \cite{schmiedeberg2011}).

While the effective diameter takes into account the fluctuations around the equilibrium-like average distance between soft spheres, the dynamical contribution of the crossing of energy barriers as studied in \cite{maiti2018} are not considered. Note that the crossing of barriers slows down the relaxation process. In the following we show that the systematic deviations described in \cite{schmiedeberg2011} can be explained in terms of temperature-dependence of the glass transition packing fraction that we have discussed in section \ref{sec:res_1}.

The data of \cite{schmiedeberg2011} is obtained for a monodisperse system at packing fractions where within the time of simulations no crystallization can be observed. As a consequence, the considered packing fractions cannot be close to the glass transition packing fraction. However, the glass transition packing fraction can be estimated, e.g., by fitting Vogel-Fulcher-Tammann functions 
\begin{eqnarray}
\nonumber
\tau(\phi,T)(p\sigma/m)^{1/2}&=A(T)\cdot\\
&\cdot\exp\left[-\frac{B(T)}{\left(\phi_{g}(T)-\phi_{eff}(T)\right)^{\delta}}\right]
\label{eq:VFT}
\end{eqnarray}
to the relaxation time as function of the packing fraction for given temperatures leading to the glass transition packing fraction $\phi_g(T)$ at that temperature. Note that there are different ways how an extrapolated glass transition packing fraction can be determined. Concerning the Vogel-Fulcher-Tammann functions different exponents $\delta$ can be employed. While we cannot decide on which $\delta$ is the best choice, we show in the following, that the obtained glass transition packing fractions in all considered cases agree to the scaling proposed in section \ref{sec:res_1}. 

First we fit Vogel-Fulcher-Tammann functions as in Eq.~(\ref{eq:VFT}) with $\delta=1$ to the data for temperatures within certain intervals (i.e., $T/\epsilon\in [10.0^{m},10.0^{m+1}]$ for $m=-9,-8,-7,-6,-5,-4,-3,-2$) in order to determine how $A(T)$, $B(T)$, and $\phi_{g}(T)$ depend on the temperature $T$. We find that within the considered temperature range $A(T)$ and $B(T)$ can be well fitted by linear functions of $\log(T/\epsilon)$, i.e., $A(T)\approx 0.16\log(T/\epsilon)+0.57$ and $B(T)\approx -0.071\log(T/\epsilon)-0.039$. Furthermore, $\phi_{g}(T)$ can be fitted by a power law: $\phi_{g}(0)-\phi_{g}(T)\propto T^\beta$ with $\beta=0.177 \pm 0.030$ and $\phi_{g}(0)=0.72\pm 0.01$ (see figure \ref{fig:fits_ges}(b)). Note that we fitted all data from within a whole temperature interval which might lead to mistakes. Therefore, in the following we extend our analysis by performing more extensive fits.

We now employ fits according to
\begin{eqnarray}
\nonumber
  \tau(\phi,T)(p\sigma/m&)^{1/2}=\left(A_1\log(T/\epsilon)+A_2\right)\cdot\\
  &\cdot \exp\left[-\frac{\left(B_1\log(T/\epsilon)+B_2\right)}{\left(\phi_{g}(0)-cT^\beta-\phi_{eff}(T)\right)^{\delta}}\right]
  \label{eq:VFT-fit}
\end{eqnarray}
with fitting constants $A_1$, $A_2$, $B_1$, $B_2$, $c$, $\phi_{g}(0)$, and $\beta$. For $\delta$ we consider either $\delta=1$ (cf. figure \ref{fig:fits_ges}(c)), $\delta=2$ (cf. figure \ref{fig:fits_ges}(d)), or we also use $\delta$ as an fit parameter (cf. figure \ref{fig:fits_ges}(e)). Note that due to the large number of fitting parameters the fitting process is not stable for all combinations of starting values. Therefore, for $A_1$, $A_2$, $B_1$, $B_2$ we employ the values that we have obtained by the method described in the previous paragraph as starting values. Our fits lead to the following results: for $\delta=1$: $\phi_{g}(0) = 0.663\pm 0.007$ and $\beta= 0.215\pm 0.009$, for $\delta=2$: $\phi_{g}(0) =0.690\pm 0.008$ and $\beta=0.208\pm 0.009$, for fitted $\delta$: $\delta=1.51\pm 0.22$, $\phi_{g}(0) = 0.677\pm 0.010$ and $\beta=0.206\pm 0.009$.

Therefore, in all cases the additional correction of $\phi_{g}(T)$ seems to be well-described by a power law with an exponent $\beta\approx 0.2$ as expected from our analysis presented in section \ref{sec:res_1}. Note that while the temperature-dependence of our fitting curves relies on the large temperature-range of the data, due to the limited range in packing fraction we cannot deduce from our fits which exponent $\delta$ or which other details of the Vogel-Fulcher-Tammann curves as functions of the packing fraction would be best to describe glassy dynamics.

\subsection{\label{sec:res_3}Scaling properties close to the glass transition}

In order to test our scaling approach for simulation data close to the glass transition, we have a closer look on the data of of Berthier and Witten \cite{BerthierandWitten,BerthierandWitten1}, who demonstrated that the relaxation times of soft spheres can be quite well collapsed by suitable rescaling of the packing fraction and the temperature onto one universal function below the glass transition density and one function above the glass transition density. Here we show that the collapse of the data can be improved by employing the scaling relation that we propose in this article. To be specific, we show that with our approach it is not necessary to choose an exponent $\mu$ that differs from the theoretically expected value $\mu=1$.

\begin{figure}[htb]
\centering
\includegraphics[width=\linewidth]{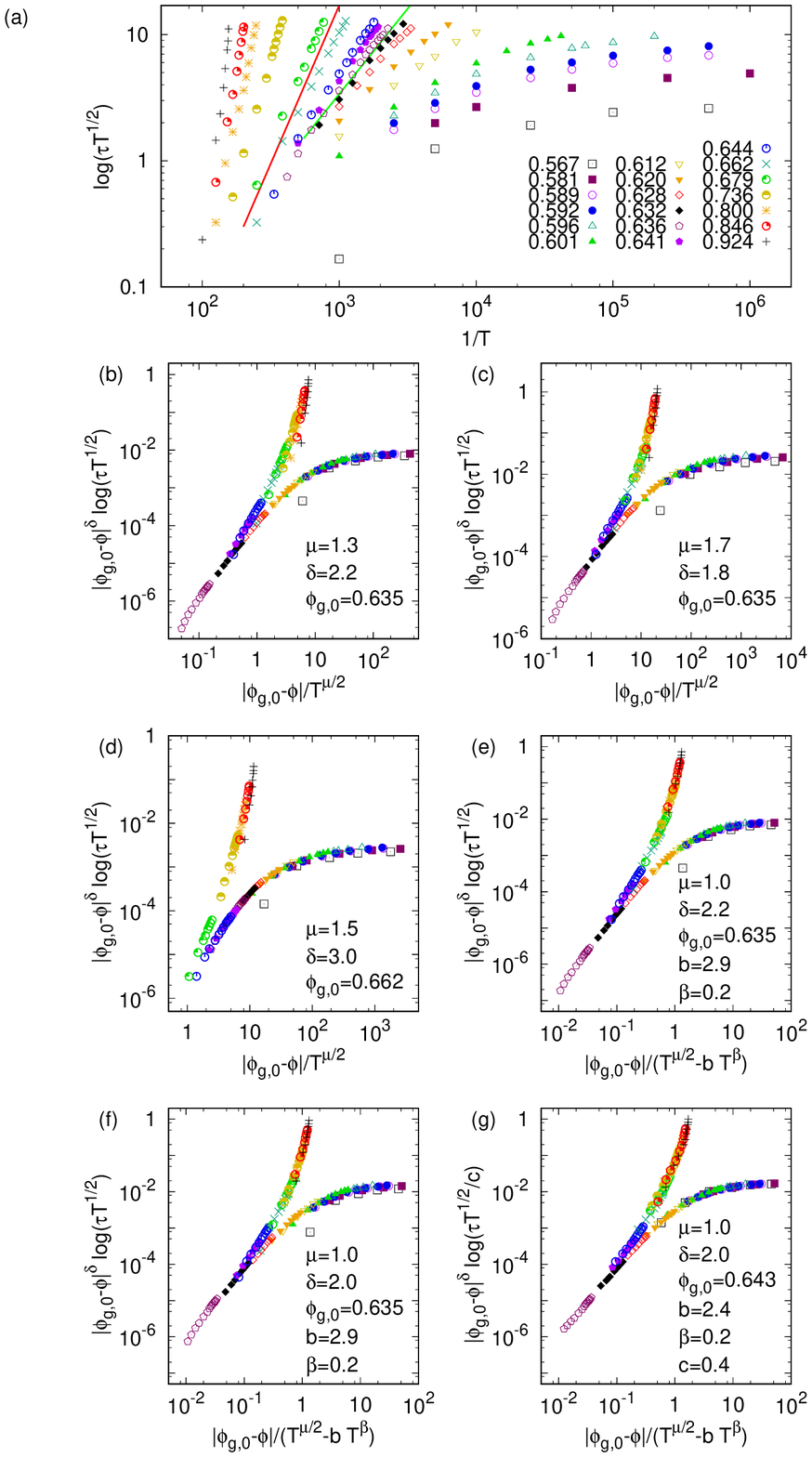}
\caption{Rescaled versions of the data from \cite{BerthierandWitten}. (a) Unscaled data in a log-log-plot such that the transition between convex and concave curves can be observed: To the right of the red line all curves are concave and therefore compatible with the lower branch in a collapsed version. To the left of the red line it is hard to tell whether the curves are concave or convex. The green line indicates the transition reported in \cite{BerthierandWitten,BerthierandWitten1}. (b) Scaling as proposed by Berthier and Witten where we use $\left(\phi_{g,0}-\phi\right)/T^{\mu}$ instead of $\left(\phi_{g,0}-\phi\right)^{2/\mu}/T$ on the $x$-axis. (c) Different values of $\mu$ and $\delta$ can lead to a similar collapse. (d) Even a different glass transition packing fraction is possible. (e) The scaling as proposed in this article for $\delta=2.2$. (f) Our scaling for $\delta=2.0$. Note, in (e,f) only a few points with $\phi=0.567$ (squares) do not yet fit. (g) An additional logarithmic correction (factor $1/c$ in the logarithm on the $y$-axis) can further improve the collapse of the lower branch if $\phi_{g,0}$ is adjusted as well.}
\label{fig:fits_berthier}
\end{figure}

As outlined in the introduction, according to Berthier and Witten \cite{BerthierandWitten,BerthierandWitten1} the best collapse is obtained by assuming an effective packing fraction $\phi_{eff}=\phi-a T^{\mu/2}$ with $\mu=1.3$, which is intermediate between $\mu=1$ as expected for distinct overlaps between harmonic spheres (see also, e.g., \cite{zhangetal}) and $\mu=3/2$ as claimed for ballistic, dilute particles\cite{BerthierandWitten}. Furthermore, the exponent of the Vogel-Fulcher-Tammann function is set to $\delta=2.2\pm 0.2$ and the best fit is claimed to occur for a glass transition at packing fraction $\phi_{g,0}=0.635\pm 0.005\pm$ \cite{BerthierandWitten,BerthierandWitten1}. The collapse is obtained by plotting $\left|\phi_{g,0}-\phi\right|^{\delta}\log{\left(\sqrt{T}\tau_{\alpha}\right)}$ as a function of $\left(\phi_{g,0}-\phi\right)^{2/\mu}/T$ \cite{BerthierandWitten,BerthierandWitten1}. The collapse works well for packing fractions close to $\phi_{g,0}$ and for small temperature, but essential deviations occur for larger $\left|\phi_{g,0}-\phi\right|$ or larger $T$. We will show in the following that better collapses can be obtained (even with less free fitting parameters). Note that the limits of the scaling approach of Berthier and Witten in \cite{BerthierandWitten} can be easily seen by considering the curvature in log-log-plots of the curves of $\log{\left(\sqrt{T}\tau_{\alpha}\right)}$ as function of $1/T$ with constant packing fraction: According to the approach of Berthier and Witten all curves below $\phi_{g,0}$ should be concave in a log-log-plot while all curves above $\phi_{g,0}$ should be convex such that they can be collapsed onto the universal scaling functions. This is because the rescaling according to Berthier and Witten corresponds to affine transformations of the curves in the log-log-plot such that concave curves are always mapped on concave curves and convex curves always on concave ones. From looking at $\log{\left(\sqrt{T}\tau_{\alpha}\right)}$ as function of $1/T$  in a log-log-plot one therefore would expect the glass transition to occur at a packing fraction larger than 0.662 where there is a transition from concave curves to maybe convex curves (see red line in figure \ref{fig:fits_berthier}(a) where we replot the data extracted from figure 2(a) of \cite{BerthierandWitten} in the described log-log-representation). To be specific, while for packing fractions larger than 0.662 it is hard to tell whether the curves are convex or concave, below or at 0.662 all curves are concave, i.e., they cannot be shifted such that they suit the convex scaling function of the upper branch as claimed in \cite{BerthierandWitten,BerthierandWitten1}. For comparison, the transition reported in \cite{BerthierandWitten,BerthierandWitten1} is indicated by a green line in \ref{fig:fits_berthier}(a). All curves to the left of that line in \cite{BerthierandWitten,BerthierandWitten1} are displaced such that they somehow arrange on a convex curve though some of the individual curves clearly are concave.

Note that the rescaling of Berthier and Witten can also be achieved by plotting $\left|\phi_{g,0}-\phi\right|^{\delta}\log{\left(\sqrt{T}\tau\right)}$ as a function of $\left(\phi_{g,0}-\phi\right)/T^{\mu}$. In figure \ref{fig:fits_berthier}(b) we use this representation for the rescaling proposed by Berthier and Witten \cite{BerthierandWitten,BerthierandWitten1}. Unlike Berthier and Witten in \cite{BerthierandWitten,BerthierandWitten1} we plot all data points and not just those that collapse onto the scaling functions. We want to point out that similar collapses can be achieved for various choices of parameters, e.g. $\delta$ can also be smaller if $\mu$ is increased (see figure \ref{fig:fits_berthier}(c)). As a consequence, the parameters $\mu$, $\delta$, and $\phi_{g,0}$ are not that well determined as Berthier and Witten suggest. Concerning $\phi_{g,0}$ we show in \ref{fig:fits_berthier}(d) that a similar collapse can be achieved for a glass transition packing fraction of 0.662 that is expected by considering the transition between concave and maybe convex curves as explained in the previous paragraph.

Now we want to demonstrate that the data of Berthier and Witten agrees with the scaling that we propose in this article. Instead of rescaling the $x$-axis by $T^{-\mu/2}$ with $\mu=1.3$ as chosen in \cite{BerthierandWitten,BerthierandWitten1} we choose the scaling of section \ref{sec:res_1}, i.e, we rescale it by $(T^{\mu/2}-bT^{\beta})^{-1}$ with $\mu=1$, $\beta=0.2$, and a parameter $b$ that we chose such to obtain the best fit. In principle other values for $\mu$ or $\beta$ could be chosen. We employ $\mu=1$ such that the effective packing fraction scales as in \cite{zhangetal} and $\beta=0.2$ in agreement to \cite{maiti2018} as explained in section \ref{sec:res_1}.  For the glass transition packing fraction we use $\phi_{g,0}=0.635$ as in \cite{BerthierandWitten,BerthierandWitten1}. The $y$-axis is left unchanged in order to resemble the Vogel-Fulcher-Tammann behavior. In figure \ref{fig:fits_berthier}(e) we employ an exponent $\delta=2.2$ as in \cite{BerthierandWitten,BerthierandWitten1}, while in figure \ref{fig:fits_berthier}(f) we use $\delta=2$. In both cases the collapse of the data, especially concerning the upper branch, is superior to the collapse of Berthier and Witten and by further varying $\delta$ (not shown) we find that $\delta=2$ seems to be the best choice of $\delta$. Note that we can also use other values for $\phi_{g,0}$, but the best collapses seem to be reached for $\phi_{g,0}$ between $0.62$ and $0.64$.

Note that in contrast to the scaling proposed by Berthier and Witten, our way of rescaling is not an affine transformation and therefore one can map a concave curve onto a convex curve such that the collapse of the upper branch can be composed of only convex curves. The collapse of the lower branch can even be further improved if the logarithmic correction due to the prefactor of the assumed Vogel-Fulcher-Tammann extrapolation function is not neglected as in  \cite{BerthierandWitten,BerthierandWitten1} such that there is an additional fit parameter $1/c$ within the logarithm on the $y$-axis. To be specific $\left|\phi_{g,0}-\phi\right|^{\delta}\log{\left(\sqrt{T}\tau_{\alpha}/c\right)}$ can be used on the $y$-axis (see, e.g., figure \ref{fig:fits_berthier}(g)). In this case the best results are obtained for transition packing fractions $\phi_{g,0}$ between $0.64$ and $0.65$. As a consequence, a different transition packing fraction might be obtained if the collapse is further improved. Note that the data in \cite{BerthierandWitten,BerthierandWitten1} was probably obtained in a way that simulations at larger packing fractions might have used the results of simulations at lower packing fractions as an input. As a consequence, in principle the ageing times for different data points might be different. Since the transition packing fraction is expected to increase with increasing ageing time \cite{berthier-prx,maiti2018} this might slightly influence the analysis. Note that today much better equilibration methods are available \cite{berthier-prx}. However, we have no indication that the data in \cite{BerthierandWitten,BerthierandWitten1} is not equilibrated sufficiently for our analysis.

\section{\label{sec:conclusions}Conclusions}

We have shown that the dynamics of soft spheres can be explained by an effective distance to the glass transition packing fraction, which depends on the temperature. One contribution due to fluctuations around mean overlaps between the soft spheres can be denoted by an effective packing fraction that can be estimated, e.g., by employing the Andersen-Weeks-Chandler method \cite{andersen71} or a correction proportional to $T^{1/2}$ \cite{zhangetal} close to the glass transition. In this article we have shown that the dynamics of soft spheres can be even better characterized if one considers an additional contribution. As discovered in \cite{maiti2018} the glass transition packing fraction decreases with an increasing probability of barrier crossing events. Here we showed that this decrease can be described by a power law $T^{0.2}$. We tested whether this decrease of the glass transition packing fraction agrees with the simulation data of \cite{schmiedeberg2011} for the dynamics of monodisperse soft spheres at packing fractions mainly between 0.35 and 0.55 as well as with the data by Berthier and Witten \cite{BerthierandWitten,BerthierandWitten1} for a bidisperse soft sphere system close to the glass transition. We have demonstrated that by considering both of the mentioned mechanisms, the data can be described better than with previously employed scaling approaches. 

Our results show that the data is in agreement with the proposed scaling of equation (\ref{eq:scaling}), which is based on theoretical considerations, i.e., $\mu=1$ as predicted from the overlap of two harmonic spheres \cite{zhangetal,BerthierandWitten,BerthierandWitten1} and $\beta=0.2$ in agreement with the scaling of the simulation results in \cite{maiti2018}. If these exponents are fixed, we can properly fit or collapse all data. Note that fits or rescaling methods where all exponents are varied as fitting parameters usually do not lead to unique sets of parameters because there are too many free fitting parameters. Therefore, while for the data of \cite{schmiedeberg2011} we obtained $\beta\approx 0.2$ by fits, we do not want to claim that we can uniquely extract all mentioned scaling exponents from the data from \cite{BerthierandWitten,BerthierandWitten1}. In a similar fashion, the respective glass transition packing fraction $\phi_{g,0}$ in the limit of zero temperature might depend on the employed extrapolation scheme, e.g., it depends on the choice of the exponent $\delta$ of the employed Vogel-Fulcher-Tammann function. Furthermore (and more important), $\phi_{g,0}$ depends on the protocol \cite{BerthierandBiroli,berthier-prx,maiti2018} and maybe also on the system size \cite{maiti2018}. Nevertheless, our scaling approach corrects for the systematic deviations that have already been reported in  \cite{schmiedeberg2011} and it can explain why concave curves can be mapped onto a convex branch in the scaling approach of  \cite{BerthierandWitten,BerthierandWitten1}.

According to \cite{medina2011,medina2012,medina2013} there is the dynamics of soft particles that can be mapped onto the dynamics of hard spheres and there is the dynamics of ultrasoft particle that might be significantly different. The difference is attributed to the occurrence of significant overlaps \cite{medina2011,medina2012,medina2013}. It would be interesting to test whether the differences are connected to the shift of the glass transition packing fraction due to the crossing of energy barriers that is studied here. Note that the Andersen-Weeks-Chandler method is also limited by the occurrence of three particles that overlap at the same time. However, it might be interesting whether more refined methods to determine the effective packing fraction can be found. Note that for very large packing fraction, reentrant glass transitions \cite{berthier10,schmiedeberg13,miyazaki16} are observed thought they can be rescaled onto a monotonic behavior \cite{schmiedeberg13}, which might facilitate to find approaches to determine effective packing fractions even at large packing fractions.

The dynamics of soft spheres can also be successfully described as cage-jumping motion \cite{pastore2016}. In addition, the dynamics in systems with other interactions might also be interesting for future studies, especially if there are addition attractions such that gelation occurs. In gel networks multiple relaxation processes on different timescales take place, e.g. the ageing of a gel network is related to a spatial directed percolation \cite{kohl}, but in principle there can be many other relaxation processes \cite{gel}. Furthermore, active particles can also lead to glassy dynamics at large densities \cite{henkes,berthieractive} and it is of large interest in ongoing research to what extend the changes of the dynamics due to an increase of activity in an active system can effectively be compared to an increase of the temperature in a passive system.

When the decrease of the glass transition packing fraction as a function of the temperature is determined, it is assumed that the crossing of barriers is a rare event and the singular rearrangements hardly contribute to the mean rearrangement dynamics \cite{maiti2018}. At such small temperatures the glass transition packing fraction can be defined by a weak ergodicity breaking \cite{maiti2018}. However, at larger temperature the rearrangement due to the crossing of energy barriers obviously cannot be neglected. At intermediate temperatures such rearrangements correspond to the ageing dynamics of a glass. However, if for a further increased temperature the timescale of ageing becomes similar to the timescale of the glassy dynamics and it is a matter of an arbitrarily chosen timescale where the glass transition packing fraction is located. 

Future works might explore how the scaling approach that we propose here is related to the rescaling that can be used for Roskilde liquids, whose dynamics only depend on one control parameter (see, e.g., \cite{Dyre}). The systems considered here - soft harmonic spheres and hard spheres - are not Roskilde liquids, but it is at least expected that their dynamics can be approximately described by one control parameter \cite{Dyre2}. In our case this control parameter is related to the distance of the effective packing fraction $\phi_{eff}(T)$ to the temperature-dependent glass transition packing fraction $\phi_g(T)$.

Finally, the effective packing fraction based on the fluctuations around the mean overlap is the same that also describes the mapping of the structural properties of soft spheres onto the structures of hard spheres. However, the decrease of the glass transition packing fraction that as we have shown is essential for a good mapping of the dynamics does not have to be considered for the structural mappings. In other words, the crossing of energy barriers changes the dynamics but not the structure of a glassy soft sphere system. This might lead to a deeper insight into how structure and dynamics of a system close to the glass transition are related.

\ack 
The project was supported by the Deutsche Forschungsgemeinschaft (Grant No. Schm 2657/3-1). We gratefully acknowledge the computer resources and support provided by the Erlangen Regional Computing Center (RRZE).

\end{document}